\documentclass[11pt,a4paper]{article}

\usepackage[a4paper,text={450pt,650pt},centering]{geometry}
\usepackage{hyperref}
\usepackage{mathrsfs}
\usepackage{bbm}
\usepackage{amsmath,amssymb,placeins}
\usepackage{verbatim}
\usepackage{stmaryrd,xcolor}
\usepackage{amsfonts,graphicx,float,ulem}

\renewcommand{\arraystretch}{0.8}

\font\tenfrak=eufm10  
\font\sevenfrak=eufm7 \font\fivefrak=eufm5
\newfam\frakfam
\textfont\frakfam=\tenfrak \scriptfont\frakfam=\sevenfrak
\scriptscriptfont\frakfam=\fivefrak

\DeclareMathOperator{\Tr}{Tr}

\begin{document}


\begin{titlepage}

\begin{centering}

\vspace{1in}

{\Large {\bf The Spectrum of the Baryon Masses in a Self-consistent SU(3) Quantum Skyrme Model}}

\vspace{.5in}

{\large D. Jur\v{c}iukonis${}^{a,1}$, E. Norvai\v{s}as${}^{a,2}$ and V. Regelskis${}^{a,b,3}$}\\
\vspace{.3 in}
${}^{a}${\textit{Institute of Theoretical Physics and Astronomy, Vilnius University,\\Go\v{s}tauto 12, Vilnius 01108, Lithuania}}\\
${}^{b}${\textit{Department of Mathematics, University of York,\\Heslington, York YO10 5DD, UK}}\\

\footnotetext[1]{{\tt darius.jurciukonis@tfai.vu.lt,}\quad
${}^{2}${\tt egidijus.norvaisas@tfai.vu.lt,} ${}^{3}${\tt
vr509@york.ac.uk}\quad } \vspace{.5in}

\vspace{.1in}

\end{centering}

\begin{abstract}
The semiclassical SU(3) Skyrme model is traditionally considered as describing a rigid quantum rotator with the profile function being fixed by the classical solution of the corresponding SU(2) Skyrme model. In contrast, we go beyond the classical profile function by quantizing the SU(3) Skyrme model canonically. The quantization of the model is performed in terms of the collective coordinate formalism and leads to the establishment of purely quantum corrections of the model. These new corrections are of fundamental importance. They are crucial in obtaining stable quantum solitons of the quantum SU(3) Skyrme model, thus making the model self-consistent and not dependent on the classical solution of the SU(2) case. We show that such a treatment of the model leads to a family of stable quantum solitons that describe the baryon octet and decuplet and reproduce their masses in a qualitative agreement with the empirical values.
\end{abstract}

\end{titlepage}


\section{Introduction}

The Skyrme model is a nonlinear field theory having localized solutions, the so-called skyrmions, that are of finite energy and are characterized by a topological charge. It is an effective theory of low-energy QCD in the limit of a large number of colours, thus it describes baryons in a weakly coupled phase as was initially argued by \cite{Skyrme61,Skyrme62,Adkins}. Indeed, the semiclassical quantization of the model has proven to be successful in describing the phenomenological properties of the baryons in the low-energy region. 

The SU(2) Skyrme model was originally defined to describe a
unitary field $U(\mathbf{x},t)$ in a fundamental representation of
the SU(2) group with a natural boundary condition $U\rightarrow
\mathbbm{1}$ at the spatial infinity, $\left\vert
\mathbf{x}\right\vert \rightarrow \infty $. This implies that the
unitary field represents a topological map $S^{3}\rightarrow
S^{3}$ with an integer-valued winding number classifying the
solitonic sectors of the model. This topological charge was
interpreted as the baryon number.

The model has been directly generalized to the case of the SU(3)
group and subsequently to the general case of the SU(N) groups
\cite{Walliser}. Both SU(2) and SU(3) versions of the model have
been canonically quantized using the collective coordinate
formalism in \cite{Fujii87,Fujii88}. It was shown that the
procedure of the canonical quantization leads to the appearance of
new terms in the explicit form of the Lagrangian of the model
that are interpreted as the quantum corrections to the mass of the
skyrmion (`quantum mass corrections'). These quantum
corrections restore the stability of the solitons that
is lost in the semiclassical approach. The instability in the 
semiclassically treated SU(2) model was shown in \cite{Bander, Braaten}. 
The method of the canonical quantization has been
subsequently generalized in \cite{Acus98, Jurciukonis} to the
cases when the field $U(\mathbf{x,}t)$ belongs to a general
representation of the SU(2) and SU(3) groups and the stability of
the solitons was explicitly shown. Interestingly, it
appeared that the aforementioned quantum corrections are
representation-dependent.

The semiclassical quantization of the SU(3) Skyrme model has
several shortcomings. For example it leads to a spectrum of masses
of the baryon octet and decuplet and some physical
characteristics of these that are not in a close agreement with
the values observed experimentally. One of the reasons for this
disagreement is that the semiclassically-treated SU(3) model does not
possess stable (semiclassical) solitons. Henceforth the classical
solution of the SU(2) model (classical profile function) or some
modification of it is used instead (see e.g.\ the overview
\cite{Weigel}). An alternative way to overcome these problems is
to consider the bound-state approach to the Skyrme model (see
e.g.\ \cite{CHK,NR}).

The purpose of the present paper is to show that the quantum mass corrections of the skyrmion that appear in the canonically quantized model are essential in ensuring the stability of the quantum solitons of the SU(3) model and realize Skyrme's original conjecture that `the mass (of the meson) may arise as a self-consistent quantal effect. This point will not be followed here, but when, for calculation purposes, we want to allow phenomenologically for a finite mass this will be done by adding to $L$ a term (proportional to $m^2_{\pi}$)' \cite{Skyrme62}. We find the stable quantum solitons by varying the complete quantum energy functional with the SU(3) octet or decuplet quantum numbers and then solving it numerically. The stability is ensured by iterative calculations.

Even though the SU(3) symmetry is not an exact flavour symmetry, by properly choosing the parameters of the model we obtain a baryon mass spectrum that is very close to the experimental one. We also focus on the influence of the Wess-Zumino-Witten (WZW), symmetry breaking (SB) and the quantum mass correction terms to the baryon masses and stability of the solitons.

The paper is organized as follows. A brief description of the SU(3) model is given in the section 2 below. In section 3 we construct the quantum Skyrme model \textit{ab initio} using the collective coordinate formalism. In section 4 the quantum energy functional is derived together with the asymptotic expression of its variation. Section 5 contains the numerical calculations of the mass spectrum of the baryons. Sections 6 accommodates the discussion and the concluding remarks.


\section{The setup of the model}

The Skyrme model is defined by the chirally symmetric Lagrangian density
\begin{align}
\bigskip \mathcal{L}_{\text{Sk}} &= -\frac{f_{\pi }^{2}}{4}\Tr\left\{\mathbf{R}_{\mu }
\mathbf{R}^{\mu }\right\}+\frac{1}{32
e^{2}}\Tr\left\{[\mathbf{R}_{\mu }, \mathbf{R}_{\nu
}][\mathbf{R}^{\mu },\mathbf{R}^{\nu }]\right\}, \label{G2}
\end{align}
where the right chiral current is defined as $\mathbf{R}_{\mu}=\left(\partial _{\mu }U\right) U^{\dagger }$. The pion decay constant $f_{\pi }$ and the dimensionless parameter $e$ are the only parameters of the model.

The main ingredient of the model is the unitary field $U:=U(\mathbf{x},t)$ that in addition to the
fundamental representation $(1,0)$ may also be defined for a general irreducible representation (irrep) $(\lambda ,\mu )$ of the SU(3) group. Then the basis states of the irrep $(\lambda ,\mu )$ are labelled by the parameters $(z,j,m)$ that are related to the hypercharge as $y=\frac{2}{3}(\mu -\lambda)-2z$ (see \cite{Jurciukonis} for the details). In such a way the classical SU(2) solitonic solution of the hedgehog type, that is defined by the canonical $SU(2) \hookrightarrow SU(3)$ embedding, takes the following form,
\begin{equation}
\exp \big( i(\mathbf{\sigma }\cdot \hat{x})F(r) \big) \hookrightarrow U_{0} \big(
\hat{x},F(r) \big) = \exp \left( 2i\Big( J_{(0,1,\cdot )}^{(1,1)}\cdot
\hat{x}\Big) F(r) \right) ,  \label{G14}
\end{equation}
where $F(r)$ is the soliton profile function, $\mathbf{\sigma }$ are Pauli matrices and $\hat{x}$ is the unit vector. The generators $J_{(0,1,\cdot )}^{(1,1)}$ represent the SU(2) subset of the SU(3) algebra generators $J_{(Z,I,M)}^{(1,1)}$. The superscript denotes that they are tensors of the adjoint representation $(1,1)$ and thus can be expressed in terms of the Gell-Mann generators (we again refer the reader to \cite{Jurciukonis} for the details).

As was shown in \cite{Jurciukonis}, the Lagrangian of the model depends on the irrep the unitary field $U(\mathbf{x},t)$ was defined for. Interestingly the dependence on the irrep appears as an overall factor of the Lagrangian \eqref{G2} and is expressed in terms of the dimension of the chosen irrep and the eigenvalue of the quadratic Casimir operator of the SU(3) algebra. Likewise the baryon number of the model includes the same overall factor. Thus the Lagrangian of the model may be normalized in such way that at the classical level it is irrep--independent. However this is not the case at the quantum level. The canonical quantization of the model leads to a quantum mass correction that is representation dependent. Furthermore, the Wess-Zumino-Witten and the symmetry breaking (mass) terms depend essentially on the chosen irrep \cite{Jurciukonis}. In this work we shall restrict to the fundamental representation of the SU(3) group only. Nevertheless we shall be using the general formalism of \cite{Jurciukonis}, hence the generalization to the higher reps is straightforward.

The hedgehog ansatz \eqref{G14} reduces the Lagrangian density \eqref{G2} to the following simple form,
\begin{align}
\mathcal{L}_{\text{cl}}[F(r)] =
-\mathcal{M}_{\text{cl}}(F(r)) = -\bigg\{{\frac{f_{\pi }^{2}}{2}}\Bigl(F^{^{\prime }2}+{\frac{2}{r^{2}}}\sin ^{2}\!F\Bigr) + {\frac{1}{2e^{2}}}{\frac{\sin
^{2}\!F}{r^{2}}}\Bigl(2F^{^{\prime }2}+{\frac{\sin
^{2}\!F}{r^{2}}}\Bigr)\bigg\},  \label{G17}
\end{align}
which represents the mass of the classical spherically symmetric soliton. Variation of this expression leads to the differential equation for the profile function $F(r)$ with topological boundary conditions $F(0)=\pi $ and $F(\infty)=0$.

For completeness we also give the explicit expressions of the symmetry breaking and the WZW terms. The SU(3) chiral symmetry breaking term is defined as \cite{Jurciukonis}
\begin{equation}
\mathcal{L}_{\text{SB}} = -{\mathcal{M}}_{\text{SB}} = \frac{f_{\pi
}^{2}}{4}\left[ m_{0}^{2}\text{ Tr}\left\{ U+U^{\dagger
}-2\cdot\mathbbm{1} \right\} +2m_{8}^{2}\text{ Tr}\left\{ \big(
U+U^{\dagger }\big) J_{(0,0,0)}^{(1,1)}\right\} \,\right]
.  \label{G7}
\end{equation}
The WZW action is given as an integral over the five dimensional
manifold $M^{5}$ the boundary of which is the compactified
spacetime, $\partial M^{5}=M^{4}=S^{3}\times S^{1}$. The standard
form for this term is
\begin{equation}
S_{\text{WZ}}(U)=-\frac{iN_{c}}{240\pi
^{2}}\int\limits_{M^{5}}\mathrm{d}^{5}x\text{\thinspace }\epsilon
^{\mu \nu \lambda \rho \sigma }\Tr\left\{\mathbf{R}_{\mu
}\mathbf{R}_{\nu }\mathbf{R}_{\lambda }\mathbf{R}_{\rho
}\mathbf{R}_{\sigma }\right\}, \label{C1}
\end{equation}
where $N_{c}$ is the number of colours.


\section{Canonical quantization}

The canonical quantization of the model is performed in terms of the collective coordinates. This approach allows the quantum unitary field to be cast in a factorisable form with the spatial and temporal dependent parts of the field being explicitly separated,
\begin{equation}
U\big(\hat{x},F(r),\mathbf{q}(t)\big) = D^{(1,0)}(\mathbf{q}(t)) \, U_{0}\left(
\hat{x},F(r)\right) D^{\dagger (1,0)}(\mathbf{q}(t))\,. \label{B1}
\end{equation}
Here $D^{(1,0)}(\mathbf{q}(t))$ is a Wigner $D$-matrix and is defined on the seven-dimensional homogeneous space SU(3)/U(1), which is specified by the seven real, independent parameters $q^{\alpha}(t)$, `the \nolinebreak collec\-tive coordinates'. The ansatz \eqref{B1} may be effectively understood as the rotation of the field $U_0$ in the quantum internal space parametrized by the collective coordinates $\mathbf{q}(t)$.

The Lagrangian (\ref{G2}) is considered quantum mechanically \textit{ab initio}. Thus the collective coordinates $q^\alpha$ and the conjugate momenta $p_\beta$ are required to satisfy the canonical commutation relations $\left[ p_{\beta},q^{\alpha}\right] =-i\delta _{\beta\alpha}$. On the other hand, this means that the coordinates $q^{\alpha}(t)$ and the velocities
$\tfrac{\mathrm{d}}{\mathrm{d}t}q^{\alpha}(t)=\dot{q}^{\alpha}(t)$ do not commute, rather they should satisfy the following commutation relation,
\begin{align}
\left[ \dot{q}^{\alpha},q^{\beta}\right] = -i f^{\alpha\beta}(q),  \label{B2}
\end{align}
where $f^{\alpha\beta}(q)$ is a function of $q^{\alpha}$ only and the explicit form of which will be determined by the consistency conditions below.

The time derivative is defined by employing the usual Weyl ordering,
\begin{align}
\partial _{0}G(q)=\frac{1}{2}\left\{ \dot{q}^{\alpha},\frac{\partial }{\partial q^{\alpha}}G(q)\right\},  \label{B4}
\end{align}
where $\{\,,\}$ represents the anticommutator. The ordering of the operators is fixed by the initial form of the Lagrangian (\ref{G2}). This allows us to avoid further ordering ambiguities in the case of the time derivatives (\ref{B4}). The ansatz (\ref{B1}) is then substituted into the Lagrangian (\ref{G2}) and followed by integration over the spatial coordinates. In such a way we obtain the Lagrangian cast in terms of the collective coordinates and velocities. Then the canonical momenta may be derived by restricting to the consideration of the terms of second order in velocities (this is because the terms of the first order in velocities vanish identically). Therefore the Lagrangian at the quantum level becomes
\begin{equation}
L_{\text{Sk}} = \,\frac{1}{2}\,\dot{q}^{\alpha }\,g_{\alpha
\beta }(q,F)\,\dot{q}^{\beta }+\left[(\dot{q})^{0}\mathrm{\,-\,order\,\,terms}\right] .  \label{B5}
\end{equation}
Here $g_{\alpha \beta }(q,F)$ is the metric tensor of the system and is expressed as
\begin{equation}
g_{\alpha \beta }(q,F) = -C_{\alpha }^{\prime (Z,I,M)}(q)(-1)^{Z+M} a_{I}(F) \, \delta _{Z,-Z^{\prime }}\,\delta_{I,I^{\prime }} \, \delta_{M,-M^{\prime }}\,C_{\beta }^{\prime (Z^{\prime },I^{\prime },M^{\prime })}(q)\,,  \label{B6}
\end{equation}
where $C_{\alpha }^{\prime (A)}(q)$ are functions of the coordinates
$q^k$ only and the explicit form of which depends on the chosen parametrization of the SU(3) group. However the explicit form of $C_{\alpha }^{\prime (A)}(q)$ does not appear in the calculations. The quantum moments of inertia of the soliton are given by the integrals over the
dimensionless variable $\tilde{r}=e f_{\pi }r$,
\begin{subequations}
\begin{align}
a_{\frac{1}{2}}(F) &= \frac{1}{e^{3}f_{\pi}}\tilde{a}_{\frac{1}{2}}(F)
 = \frac{1}{e^{3}f_{\pi }}2\pi \int \mathrm{d}\tilde{r}\tilde{r}^{2}\left( 1-\cos F\right)
 \left[1+\frac{1}{4}F^{\prime 2}+\frac{1}{2\tilde{r}^{2}}\sin ^{2}F\right] ,  \label{B81} \\
a_{1}(F) &= \frac{1}{e^{3}f_{\pi}}\tilde{a}_{1}(F)
 = \frac{1}{e^{3}f_{\pi }}\frac{8\pi }{3}\int \mathrm{d}\tilde{r}\tilde{r}^{2}\sin ^{2}F
  \left[ 1+F^{\prime2}+\frac{1}{r^{2}}\sin ^{2}F\right].  \label{B8}
\end{align}
\end{subequations}
Note that $a_{0}(F)=0$ and the summation in \eqref{B6} is over the basis states $(Z,I,M)$ of irrep $(1,1)$ excluding the state $(0,0,0)$. The quantum `moment of inertia' $a_1(F)$ of the SU(3) model coincides with the quantum momentum $a(F)$ of the SU(2) model. It is important to note that $a_{1}(F)$ is not equal to the mechanical momentum of inertia of the mass distributed by the classical spherically symmetric hedgehog field defined in (\ref{G17}).

The canonical momentum, which is conjugate to ${q}^{\beta }$ is defined as
\begin{equation}
p_{\beta }^{(0)}=\frac{\partial L_{Sk}}{\partial \dot{q}^{\beta
}}=\frac{1}{2}\left\{ \dot{q}^{\alpha },g_{\alpha \beta }\right\}.
\label{B9}
\end{equation}
The superscript $^{(0)}$ was introduced to denote the canonical momentum obtained from \eqref{B5}. As we will show later, the WZW term shall contribute to the final form of the canonical momentum. Next, by requiring the canonical commutation relations $\left[ p_{\beta}^{(0)},q^{\alpha }\right] =-i\delta _{\alpha \beta }$ to be satisfied, the initially undetermined commutation relations \eqref{B2} are constrained to be
\begin{align}
\left[ \dot{q}^{\alpha},q^{\beta}\right] = -i g^{\alpha \beta }(q,F)\,, \label{B2A}
\end{align}
where $g^{\alpha \beta }(q,F)$ is the inverse of \eqref{B6}. This relation allows us to determine the explicit form of the $(\dot{q})^{0}$--order terms in \eqref{B5}. Thus after substituting \eqref{B1} into \eqref{G2}, carefully manipulating the non-commutative variables and integrating over the spatial coordinates the additional quantum mass corrections are revealed \cite{Jurciukonis}. Their explicit form will be presented in the section below.

The contribution of the WZW term to the effective Lagrangian of the Skyrme model in the framework of the
collective coordinate formalism was considered in \cite{Balachan}. By plugging \eqref{B1} into \eqref{C1} and employing Stokes's theorem and performing careful calculations, the WZW term takes the following form,
\begin{equation}
L_{\text{WZ}}(q,\dot{q})=-\lambda ^{\prime }\frac{i}{2}\left\{
\dot{q}^{\alpha },C_{\alpha }^{\,\prime (0)}(q)\right\} , \label{C2}
\end{equation}
where $\lambda ^{\prime }=\tfrac{N_{c}B}{2\sqrt{3}}$ and $B$ is the baryon number.

The Lagrangian of the system with the inclusion of the WZW term becomes $L^{\prime }=L_{\text{Sk}}+L_{\text{WZ}}$. The WZW term may be considered as an external potential of the system \cite{Mazur}. Therefore it shifts the canonical momenta $p^{(0)}_{\beta }$ \eqref{B9} by
\begin{equation}
p_{\beta } = \frac{\partial L^{\prime }}{\partial \dot{q}^{\beta}} =
 \frac{1}{2}\left\{ \dot{q}^{\alpha },g_{\alpha \beta }\right\} - i\lambda ^{\prime }C_{\beta }^{\,{\prime }(0)}(q) \,.  \label{C5}
\end{equation}
The metric tensor $g_{\alpha\beta}$ and the functions $f^{\alpha\beta}$ are not modified and the canonical commutation relations are preserved.


\section{The Hamiltonian}

The Lagrangian $L^{\prime }=L_{\text{Sk}}+L_{\text{WZ}}$ effectively describes a system on a curved space with the metric $g_{\alpha \beta }(q,F)$ defined by \eqref{B6}. The Hamiltonian for such a system is obtained by employing the general method of quantization on the curved space developed by Sugano \textit{et al.} \cite{Sugano}. This ensures the consistency of the Hamiltonian with the Euler-Lagrange equations of the model.

We start by introducing seven right transformation generators
\begin{equation}
\hat{R}_{(\bar{A})}=\frac{i}{2}\left\{ p_{\alpha }+\lambda
^{\prime }iC_{\alpha }^{\prime (0)}(q),C_{(\bar{A})}^{\prime
\alpha }(q)\right\} \label{C6}
\end{equation}
that satisfy standard commutation relations of the SU(3) algebra.
Here index $\bar{A}$ denotes the set $(Z,I,M)$ excluding the case
$(0,0,0)\,$, and $C_{(\bar{A})}^{^{\prime }\alpha }(q)$ are the
reciprocal functions to $C_{\alpha }^{\prime(\bar{A})}(q)$, thus
satisfy the standard orthogonality conditions. The generators 
$\hat{R}_{(0,1,\cdot)}$ form a SU(2) subalgebra of SU(3) and may 
be interpreted as spin operators. This is because their
action on the unitary field can be realized as a spatial rotation
of the skyrmion only. Next, it is convenient to define the eight
transformation generator as $\hat{R}_{(0,0,0)} =
-\lambda^{\prime}$ or equally $Y_S=1$ in (\ref{C16}).
In a similar way eight left transformation generators may be introduced,
\begin{equation}
\hat{L}_{(B)}=\frac{1}{2}\left\{ \hat{R}_{(A)},D_{(A)(B)}^{\dagger(1,1)}(q)\right\},  \label{C9}
\end{equation}
using which the effective Hamiltonian of the model (with the constraint
$\hat{R}_{(0,0,0)}=-\lambda ^{\prime }$ included) is found to be (see \cite{Jurciukonis} for the details)
\begin{align}
H^{\prime } &= \frac{1}{2a_{\frac{1}{2}}(F)}\left((-1)^{A}\hat{L}_{(A)}\hat{L}_{(-A)}-\lambda ^{\prime 2}\right)
 + \frac{1}{2}\left(\frac{1}{a_{1}(F)}-\frac{1}{a_{\frac{1}{2}}(F)}\right)(-1)^{m}\hat{R}_{(0,1,m)}\hat{R}_{(0,1,-m)}  \notag \\
& \quad + \Delta M_{1}+\Delta M_{2}+\Delta M_{3}+M_{\text{cl}} \,, \label{C13}
\end{align}
where the following notation has been introduced:
\begin{subequations}
\begin{align}
\Delta M_{1} &=-\frac{2\pi }{a_{1}^{2}(F)}\int r^{2}\mathrm{d}r \, \sin ^{2}F\left[ f_{\pi
}^{2}+\frac{1}{2e^{2}}\left( 2F^{\prime 2}+\frac{\sin ^{2}F}{r^{2}}\right) \right], \\
\Delta M_{2} &= -\frac{\pi }{a_{\frac{1}{2}}^{2}(F)}\int r^{2}\mathrm{d}r \,(1-\cos F) \left[ f_{\pi }^{2}(2-\cos F)+\frac{1}{4e^{2}} \left( (2+\cos F)F^{\prime 2}+\frac{2\sin ^{2}F}{r^{2}}\right) \right], \\
\Delta M_{3} &= -\frac{2\pi }{a_{1}(F)a_{\frac{1}{2}}(F)}\int r^{2}\mathrm{d}r \,\sin ^{2}F\left[ f_{\pi}^{2}+\frac{1}{2e^{2}}\left( F^{\prime 2}+\frac{\sin^{2}F}{r^{2}}\right) \right].
\end{align} \label{QMC}
\end{subequations}

\noindent These negative quantum mass corrections appear because of the
non-trivial commutation relations of the quantum coordinates and
velocities \eqref{B2A} and were first derived in \cite{Fujii88}. This approach was later generalized for the field $U({\bf x},t)$ in a general representation $(\lambda,\mu)$ of SU(3) in \cite{Jurciukonis}. Equations \eqref{QMC} correspond to the fundamental representation $(1,0)$ of the general case given in \cite{Jurciukonis} and are equivalent to the ones given in \cite{Fujii88} (up to some misprints).
The kinetic part of the effective Hamiltonian is a differential operator constructed from the 
SU(3)/SU(2)--left and SU(2)-- right transformation generators, thus the
eigenstates of the model are
\begin{equation}
\textstyle \genfrac{|}{\rangle}{0pt}{}{(\Lambda, M )}{\scriptstyle
Z_T(Y_T),T,M_{T};~Z_S(Y_S),S,M_{S}}
 = \sqrt{\dim (\Lambda ,M)}\, D_{(Z_T,T,M_{T})(Z_S,S,M_{S})}^{\ast (\Lambda M)}(q)\left\vert 0\right\rangle,  \label{C16}
\end{equation}
where the quantity $D^*$ on the right-hand side is the complex conjugate matrix element of the Wigner $D$-matrix for the $(\Lambda, M)$ irrep of the SU(3) group and is expressed in terms of the quantum variables $q^{k}$. The topology of the eigenstates can be non-trivial and the quantum states contain an eighth `unphysical' quantum variable $q^{0}$.

Finally we are ready to consider the symmetry breaking term which takes the following form,
\begin{equation}
L_{\text{SB}}=-M_{\text{SB}}=4\pi f_{\pi }^{2}\text{ }\int r^{2}\mathrm{d}r(1-\cos F)\left[m_{0}^{2}-\frac{1}{\sqrt{3}}\,m_{8}^{2} \,D_{\;(0)(0)}^{\dagger(1,1)}(q)\right] ,
\label{S4}
\end{equation}
where the parameters $m_{0}^{2}$ and $m_{8}^{2}$ are considered as
the phenomenological parameters of the model. The expression
(\ref{S4}) contains the operator $D^{\dagger(1,1)}_{\;(0)(0)}(q)$, which is
a function of the quantum variables $q^{\alpha}$ and acts
non-diagonally on the states \eqref{C16}. This means that
$\left[\hat{L}_{(Z,\frac{1}{2},M)},M_{\text{SB}}\right] \neq 0$.
Therefore the physical states of the system with the symmetry breaking term included need to be calculated by diagonalizing the total Hamiltonian as it is done in the strong symmetry breaking limit, see \cite{Yabu} and \cite{Park}. However the contribution of the symmetry breaking term is minor compared with the rest of the Hamiltonian and thus may be considered as a first order perturbation. The matrix elements of the symmetry breaking operator can be expressed in terms of two SU(3) Clebsch-Gordan coefficients,
\begin{align}
& \textstyle \genfrac{\langle}{|}{0pt}{}{(\lambda^{\prime},\mu^{\prime}
)}{\scriptstyle{Y^{\prime}_T,T^{\prime},M^{\prime}_T;~
Y^{\prime}_S,S^{\prime},M^{\prime}_S}}D^{(1,1)}_{(0),(0)}(q)
\genfrac{|}{\rangle}{0pt}{}{(\lambda ,\mu
)}{\scriptstyle{Y_T,T,M_T;~
Y_S,S,M_S}} = \notag \\
& \quad = \frac{\dim(\lambda, \mu)}{\dim(\lambda^{\prime}, \mu^{\prime})}
\sum_{\gamma} \left[
\renewcommand{\arraystretch}{0.8}
\begin{array}{ccc}
\scriptstyle(\lambda,\mu) & \scriptstyle(1,1) & \scriptstyle(\lambda^{\prime},\mu^{\prime})_{\gamma} \\
\scriptstyle{Y_T,T,M_T} & \scriptstyle 0 &
\scriptstyle{Y^{\prime}_T,T^{\prime},M^{\prime}_T}
\end{array}
\right] \left[
\renewcommand{\arraystretch}{0.8}
\begin{array}{ccc}
\scriptstyle(\lambda,\mu) & \scriptstyle(1,1) & \scriptstyle(\lambda^{\prime},\mu^{\prime})_{\gamma} \\
\scriptstyle{Y_S,S,M_S} & \scriptstyle 0 &
\scriptstyle{Y^{\prime}_S,S^{\prime},M^{\prime}_S}
\end{array}
\right].
\end{align}

In the semiclassical approach the unitary field $U(\mathbf{x},t)$
can be expanded in power series around the classical vacuum
$U=\mathbbm{1}$. In such expansion the parameters of the symmetry
breaking term are obtained to be $m_{0}^{2}=\frac{1}{3}\left(
m_{\pi }^{2}+2\frac{f_{K}^{2}}{f_{\pi}^{2}}m_{K}^{2}\right)$ and
$m_{8}^{2}=\frac{2}{\sqrt{3}}\left(\frac{f_{K}^{2}}{f_{\pi
}^{2}}m_{K}^{2}-m_{\pi }^{2}\right)$ where the experimental ratio
$\frac{f_{K}}{f_{\pi }}=1.197\,$ is imposed in order to obtain the
standard mass terms of the $\pi$ and $ K$ mesons. 
However we treat the model quantum mechanically {\it ab initio} and the collective
coordinates $q$ are {\it not } small perturbations. Thus the
parameters $m_{0}^{2}$ and $m_{8}^{2}$ need to be treated as
generic parameters of the model.

Putting all the ingredients together the energy functional of the quantum skyrmion in the operational form for the states in the irrep $(\Lambda ,M) $ becomes
\begin{equation}
E(F) = \frac{C_{2}^{\text{SU(3)}}(\Lambda ,M)-\lambda
^{\prime2}}{2a_{\frac{1}{2}}(F)} + \frac{1}{2}\left(
\frac{1}{a_{1}(F)} - \frac{1}{a_{\frac{1}{2}}(F)}\right)S(S+1) +
\Delta M+M_{\text{cl}} + \left\langle M_{\text{SB}}\right\rangle,
\label{Efunct}
\end{equation}
where $\Delta M=\sum \Delta M_{k}\,$ and $\left\langle M_{\text{SB}}\right\rangle$ represents the symmetry breaking operator $M_{\text{SB}}$ sandwiched between the states \eqref{C16}. The variation of the energy functional $\frac{\delta E(F)}{\delta F}=0$ gives an integro-differential equation for the profile function $F(r)$ with the topological boundary conditions $F(0)=\pi$ and $F(\infty)=0$ imposed on top. At large distances this equation reduces to the asymptotic form
\begin{equation}
\tilde{r}^{2}F^{\prime \prime }+2\tilde{r}F^{\prime
}-(2+\tilde{m}^{2}\tilde{r}^{2})F=0\,,  \label{F5}
\end{equation}
where the dimensionless quantity $\tilde{m}^{2}$ is defined as
\begin{align}
\tilde{m}^{2} &= -e^{4}\Biggl(\frac{1}{4\tilde{a}_{\frac{1}{2}}^{2}(F)}\left(C_{2}^{\text{SU(3)}}(\Lambda ,M)-S(S+1) - \lambda ^{\prime 2}+1\right) + \frac{2\,S(S+1)+3}{3\,\tilde{a}_{1}^{2}(F)}  \notag \\
& \qquad\qquad + \frac{8\Delta \tilde{M}_{1} + 4\Delta\tilde{M}_{3}}{3\,\tilde{a}_{1}(F)} + \frac{\Delta
\tilde{M}_{3} + 2\Delta\tilde{M}_{2}}{2\,\tilde{a}_{\frac{1}{2}}(F)} + \frac{1}{\tilde{a}_{1}(F)\tilde{a}_{\frac{1}{2}}(F)}\Biggl) \,+ \left\langle\tilde{M}_{\text{SB}}\right\rangle .  \label{F6}
\end{align}
The tilded integrals $\Delta\tilde{M}_k$ and
$\tilde{M}_{\text{SB}}$ are calculated using the dimensionless
parameter $\tilde{r}=e f_{\pi}r$. The quantity $m=ef_{\pi}\tilde
m$ is interpreted as the effective asymptotic mass of the baryon.
For example, in the case of the nucleon it is
$m^{\text{as}}_N=m_{\pi}=137.7$ MeV. The corresponding asymptotic
solution of \eqref{F5} is found to be
\begin{equation}
F(\tilde{r})=k\left(
\frac{\tilde{m}^{2}}{\tilde{r}}+\frac{1}{\tilde{r}^{2}}\right)
\exp (-\tilde{m}\tilde{r}).  \label{F7}
\end{equation}
This solution is very important in ensuring the stability of the
quantum soliton. It effectively translates into the requirement
the integrals (\ref{B81}, \ref{B8}) and $\Delta M_{k}$ be
convergent. Such a requirement is satisfied only if the asymptotic
mass of the baryon $\tilde{m}^{2}>0$. This condition is only
satisfied in the presence of the negative quantum mass corrections
$\Delta M_{k}$ and symmetry breaking term $\langle
M_{\text{SB}}\rangle$ or at least one of them. However the general
(non-asymptotic) integro\-differential equation for the profile
function $F(\tilde{r})$ obtained from the variation $\frac{\delta
E(F)}{\delta F}=0$ of the SU(3) model does not have stable solutions when the quantum
mass corrections $\Delta M_{k}$ are absent. 

Finally we note that the symmetry breaking term is not necessary in 
ensuring the stability of the solitonic solution of the canonically 
quantized Skyrme model. The profile function $F(\tilde{r})$ has the 
required asymptotic exponential behavior (\ref{F7}) even in the chiral 
limit when the symmetry breaking term is absent. In such way the 
canonically quantized Skyrme model is self-consistent.


\section{Numerical results}

We want to estimate the influence of the quantum mass correction
$\Delta M$ on the stability of the quantum solitons and to
compare the mass spectrum of the baryon octet and decuplet
obtained using the semiclassical (rigid) and the quantum (soft) profile
functions.

\begin{figure}
\begin{center}
\includegraphics{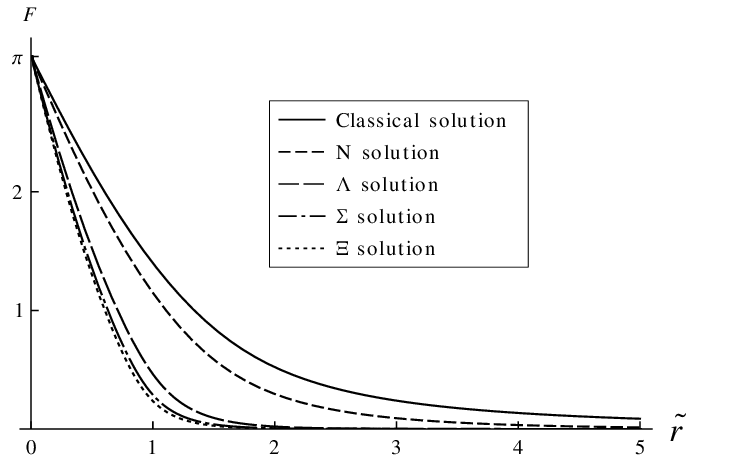}
\end{center}
\caption{The classical profile function $F(\tilde r)$ together with the quantum profile functions $F_i(\tilde r)$ of the stable quantum solitons describing the baryon octet. The quantum profile functions correspond to the calculations presented in column M$_3$ of table \ref{table1}.}
\label{picture1}
\end{figure}

Let us start by considering the semiclassical case first. The
initial step is to find the classical profile function minimizing
the energy functional of the classical SU(2) Skyrme model
\eqref{G17}. The determined profile function asymptotically decays
according to the power law $F(\tilde{r}\rightarrow \infty
)\sim\frac{1}{\tilde{r}^{2}}$ \ and respects the topological
boundary condition $F(0)=\pi $ (see figure \ref{picture1}). 
Then adding the symmetry breaking term modifies the profile 
function to be of the exponentially decaying form, $F(\tilde{r}\rightarrow \infty
)\sim\frac{1}{\tilde{r}^{2}}\,e^{-\tilde{m}_\pi \tilde{r}}$. The
next step is to choose the parametrization scheme of the model.
The SU(3) Skyrme model is parametrized by four parameters $f_{\pi
}$, $m_{0}^{2}$, $m_{8}^{2}$ and $e$. The first three parameters
are of phenomenological origin, while the last one ($e$) is a
dimensionless parameter that is usually constrained by requiring
the model to fit the experimental data. Let us name these
parameters the \textit{essential} ones as they appear in the
model explicitly. We shall also consider the following four
phenomenological parameters: the nucleon mass $m_N=939$ MeV, the
asymptotic nucleon mass $m^{\text{as}}_N=m_\pi=137.7$ MeV, the
mean nucleon isoscalar (electric) radius $\langle r^2\rangle
^{1/2}=0.78~\text{fm}$ and the mass of one of the heavier baryons
(e.g.\ $m_\Lambda$ or $m_\Sigma$) as possible input parameters
for the model. We name them the \textit{fit} parameters as they
will be used to fit the model to the experimental data. 

The results of the numerical calculations using \eqref{Efunct} and based on the classical profile function are displayed in columns M$_1$ and M$_2$ of table \ref{table1}. Here the first column displays the experimental mass spectrum of the baryon octet and decuplet (the states are not discriminated by their spin polarization). The numbers standing at the right side of the mass show the deviation {\small($\pm$\%)} of the calculated value from the experimental one. The parametrization of $m_0^2$ and $m_8^2$ for column M$_{1}$ is $m_{0}^{2}=\frac{1}{3}\left(m_{\pi
}^{2}+2\frac{f_{K}^{2}}{f_{\pi}^{2}}m_{K}^{2}\right)= 241\,032\;\text{MeV}^2$ and $m_{8}^{2}=\frac{2}{\sqrt{3}}\left(\frac{f_{K}^{2}}{f_{\pi}^{2}}m_{K}^{2}-m_{\pi}^{2}\right) = 384\,638\;\text{MeV}^2$, where $m_{\pi}=137.7$~MeV, $m_{K}=495.7$~MeV and $f_\pi = 92.2$ MeV, $\frac{f_{K}}{f_{\pi }}=1.197\,$. 

Note that the set of input parameters in both cases is different but always consists of both essential and fit parameters. The dimensionless parameter $e$ is never an input parameter and is obtained by fitting the model to the experimental data. Let us explain both choices of the input parameters in detail.

The standard choice in the semiclassical approach is to choose $f_{\pi}$, $m_{0}^{2}$, $m_{8}^{2}$ and $m_N$ as the set of input parameters describing the model (column M$_{1}$ of table \ref{table1}). However, restricting to the experimental value of $f_{\pi }$ even in the case of the SU(2) Skyrme model hardly reproduces the correct mass spectrum of the nucleon and its delta resonances \cite {Battye}. Furthermore this choice leads to a value of the mean nucleon isoscalar radius, which may be evaluated using the following expression,
\begin{equation}
\bigl\langle r^{2}\bigr\rangle = -\frac{2}{\pi e^{2}f_{\pi }^{2}}\int
\!r^{2}F^{\prime }\sin ^{2}\!F\mathrm{d}\tilde{r}\,,  \label{radius}
\end{equation}
far from the experimental one. Thus there is no particular reason to restrict to this set of input parameters and an alternative reasonable choice of the input parameters is $m_N$, $m^{\text{as}}_N$, $m_\Lambda$ and $\bigl\langle r^{2}\bigr\rangle=0.78~\text{fm}$ leading to a much better agreement with the experimental data (column M$_{2}$ of table \ref{table1}). 

\begin{table} [H]
\begin{center}
\begin{tabular}{|c|c|c|c|c|c|}
\hline \hline
MeV$_{\text{exp}}$ & $\mathbf{M_1}$ (\%) & $\mathbf{M_2}$ (\%) &  $\mathbf{M_3}$ (\%) & $\mathbf{M_4}$ (\%) & $\mathbf{m^{\text{as}}}$ \\
\hline
$N ~(939)$ & 939 (\small{Input}) & 939 (\small{Input}) & 939 (\small{Input})  & 939 (\small{Input}) & \!137.7 (\small{Input})\! \\
\hline
$\Lambda ~(1115.6)$ & 1015.0 (-9.0) & 1115.6 (\small{Input}) & 1067.0 (-4.4) & 1115.6 (\small{Input}) & 512.0 \\
\hline
$\Sigma ~(1193.1)$ & 1091.0 (-8.6) & 1292.2 (8.3) & 1193.1 (\small{Input}) & 1202.1 (0.7) & 675.6  \\
\hline
$\Xi ~(1318)$ & 1129.1 (-14.3) & 1380.5 (4.7) & 1256.3 (-4.7) & 1235.9 (-6.3) & 741.0 \\
\hline\hline
$\Delta ~(1232)$ & 1476.3 (19.8) & 1255.5 (1.9) & --- & 1373.6 (11.5) & 377.7  \\
\hline
$\Sigma^{*} ~(1385)$ & 1523.8 (10.0) & 1365.9 (-1.4) & 1461.0 (5.5) & 1468.7 (6.0) & 496.3  \\
\hline
$\Xi^{*} ~(1533.5)$ & 1571.4 (2.5) & 1476.3 (-3.7) & 1537.9 (0.3) & 1541.1 (0.5) & 590.2  \\
\hline
$\Omega^{0} ~(1672)$ & 1618.8 (-3.2) & 1586.6 (-5.1) & 1616.3 (-3.3) & 1600.7 (-4.3) & 670.2  \\
\hline\hline
$e$ & 5.7 & 3.9 & 4.5 & 3.8 & ---   \\
\hline
$f_{\pi} $ & \small{Input} & 60.9 & 92.3 & 58.3 & ---   \\
\hline
$m^2_0 $ & \small{Input} & 58 537 & 0 (\small{Input}) & 211 224 & ---  \\
\hline
$m^2_8 $ & \small{Input} & 390 361 & 546 807 & 1 273 462 & --- \\
\hline\hline
\end{tabular}
\end{center}
\caption{Baryon mass spectrum (MeV) obtained using \eqref{Efunct}. Columns M$_{1}$ and M$_{2}$ display the mass spectrum based on the classical SU(2) profile function. Column M$_{3}$ is based on the quantum profile function of the nucleon. Column M$_{4}$ is based on the individual quantum profile function for each state. 
The choice of the input parameters for each column is different and is
emphasized in the table by {\small(Input)}. The fourth input parameter for column M$_3$ is the isoscalar nucleon radius $\langle r^2\rangle ^{1/2}=0.78~\text{fm}$. The third and fourth input
parameters for columns M$_2$ and M$_4$ are the isoscalar nucleon radius and the asymptotic nucleon mass
$m^{\text{as}}_{N}=m_{\pi}=137.7$~MeV.
The last column displays the asymptotic mass spectrum of the corresponding states obtained from the
calculations of column M$_4$.} \label{table1}
\end{table}

The approach we have been considering so far is not entirely semiclassical as we have been calculating the mass spectrum with the help of the classical profile function and \eqref{Efunct} which includes the quantum mass correction $\Delta M$. However omitting this term leads to a complex value of the model parameter $e$ and thus some other method to ensure the consistency of the model needs to be employed (see e.g.\ \cite{Weigel}).

Let us turn now to the consideration of the self-consistent quantum
SU(3) Skyrme model. The main difference with respect to the
previous case is that instead of using the classical profile
function we minimize the quantum energy functional \eqref{G17} by employing recursive calculations and
thus obtain stable quantum profile functions for each state individually. The recursive calculations
are performed in the following way:

\begin{enumerate}
\item Find the classical profile function $F^{\text{\tiny{(0)}}}(\tilde{r})$ minimizing the energy functional of the classical SU(2) Skyrme model \eqref{G17} and choose the set of input parameters describing the model as discussed above.

\item Calculate the classical values of the integrals $a_1(F^{\text{\tiny{(0)}}})$, $a_{\frac{1}{2}}(F^{\text{\tiny{(0)}}})$ and $\Delta M(F^{\text{\tiny{(0)}}})$ in (\ref{Efunct}) and the (essential) model parameters by requiring the classical profile function to reproduce the physical properties of the nucleon and arbitrary heavier baryon, e.g.\ $\Lambda$.

\item Find the first approximation of the quantum profile functions $F_{N}^{\text{\tiny{(1)}}}(\tilde{r})$ and $F_{\Lambda}^{\text{\tiny{(1)}}}(\tilde{r})$ by employing the asymptotic solution (\ref{F7}) and minimizing the quantum energy functional \eqref{Efunct}, i.e.\ solving the variational equation $\frac{\delta E(F)}{\delta F}=0$ by using the classical values of the integrals $a_1(F^{\text{\tiny{(0)}}})$, $a_{\frac{1}{2}}(F^{\text{\tiny{(0)}}})$ and $\Delta M(F^{\text{\tiny{(0)}}})$, and the model parameters. Functions $F_{N}^{\text{\tiny{(1)}}}(\tilde{r})$ and $F_{\Lambda}^{\text{\tiny{(1)}}}(\tilde{r})$ are found independently as they are describe the states with different quantum numbers.\label{item3}

\item The obtained functions $F_{N}^{\text{\tiny{(1)}}}(\tilde{r})$ and $F_{\Lambda}^{\text{\tiny{(1)}}}(\tilde{r})$ are used to calculate the updated values of the integrals $a_1(F_N^{\text{\tiny{(1)}}})$, $a_{\frac{1}{2}}(F_N^{\text{\tiny{(1)}}})$, $\Delta M(F_N^{\text{\tiny{(1)}}})$, and $a_1(F_\Lambda^{\text{\tiny{(1)}}})$, $a_{\frac{1}{2}}(F_\Lambda^{\text{\tiny{(1)}}})$, $\Delta M(F_\Lambda^{\text{\tiny{(1)}}})$. The updated values of the model parameters are found by requiring the obtained profile functions to reproduce the physical properties of $N$ and $\Lambda$. Then the procedure described in item~\ref{item3} is repeated to get the second approximation of the quantum solutions $F_{N}^{\text{\tiny{(2)}}}(\tilde{r})$ and $F_{\Lambda}^{\text{\tiny{(2)}}}(\tilde{r})$.\label{item4}

\item The procedure described in item~\ref{item4} is iterated until the convergent solutions
$F_{N}(\tilde{r})$, $F_{\Lambda }(\tilde{r})$ and stable values of the integrals $a_1(F_N)$, $a_{\frac{1}{2}}(F_N)$, $\Delta M(F_N)$, and $a_1(F_\Lambda)$, $a_{\frac{1}{2}}(F_\Lambda)$, $\Delta M(F_\Lambda)$, and the model parameters are obtained.

\item The obtained model parameters are used to find the quantum profile functions for the rest of the baryons. The same iteration procedure is employed (with the model parameters fixed) until the convergent solution and stable integrals are obtained.

\end{enumerate}

In case of the semiclassical approach this procedure fails -- it does not lead to a stable soliton due to the absence of the quantum mass correction $\Delta M$, which not only contributes to the asymptotic mass of the state (\ref{F6}) which is required to be real and positive, but also plays a crucial role in  solving the variational equation $\frac{\delta E(F)}{\delta F}=0$. 

The mass spectrum of the quantum SU(3) model is presented in column M$_4$ of table \ref{table1}. The choice of input parameters is the same as for column M$_2$. Each state is described by an individual profile function obtained using steps 1-6 explained above and is displayed in figure~1. These quantum profile functions are very important as they can be used to calculate the magnetic moments and form factors of the corresponding states. The obtained mass spectrum is very close to the experimental one except for the $\Delta$ state. However delta resonances are not stable baryons; thus are not expected to be described by the model very well.

The last column of table \ref{table1} displays the asymptotic baryon mass spectrum. It reflects the mass density of the corresponding states in the asymptotic region, $r \rightarrow \infty $. 

Finally let us discuss the calculations presented in column M$_3$. The interesting fact is that such approach predicts the correct value of $f_\pi$, while the approach used for columns M$_2$ and M$_4$ leads to a value of $f_\pi$ much smaller than the experimental one. However this approach does not describe the $\Delta$ state as the corresponding integrals \eqref{QMC} diverge.


\section{Discussion}

In this work we have considered the stability of the topological solitons of the quantum SU(3) Skyrme model formulated in \cite{Jurciukonis}. The model was shown to possess a family of stable quantum solitons whose energy functionals reproduce the mass spectrum of the baryon octet and decuplet in a good agreement with the experimental results.

The semiclassical and quantum Skyrme models are essentially different models and lead to distinct
integro-differential equations for the profile function $F(\tilde{r})$. In the semiclassical approach the energy functional $E(F)$ does not receive quantum corrections \eqref{QMC} and the symmetry breaking term plays an important role in obtaining the exponentially decaying asymptotic profile function. Despite having correct asymptotic behavior, the semiclassical SU(3) Skyrme model does not support stable (quantum) solitons. Recursive solutions of the variational equation $\frac{\delta E(F)}{\delta F}=0$ {\it do not} converge and the classical profile function must be used instead. Hence the semiclassical Skyrme model is considered as describing a {\it rigid} quantum rotator because the profile function is fixed by the classical solution.

The canonical quantization of the Skyrme model leads to the appearance of the quantum mass corrections \eqref{QMC} in its energy functional. These corrections not only ensure the correct asymptotic form of the profile function even in the absence of the symmetry breaking term, but also are necessary for obtaining stable quantum solitons with fixed baryon quantum numbers. The recursive solutions of the variational equation $\frac{\delta E(F)}{\delta F}=0$ with the quantum mass corrections present {\it do} converge and lead to quantum profile functions which differ from the classical one. The difference is explicitly shown in the figure \ref{picture1} \ where the classical profile function and the quantum profile functions for the baryon octet are displayed.

Interestingly, the stability is preserved even if the Wess-Zumino-Witten and the symmetry breaking terms are not included in the model. Thus in this sense the quantum SU(3) Skyrme model is self-consistent and may be effectively understood as describing a {\it soft} quantum rotator.
Despite the model being self-consistent the symmetry breaking term is necessary as it is responsible for the discrimination of the solutions with different hypercharges. Thus it must be included into the model in order to obtain physically reasonable results.

The quantum approach to the SU(3) Skyrme model not only makes the
quantum solitons stable, but also adjusts the model to fit
better to the experimental results. Our numerical calculations of
the mass spectrum of the octet and the decuplet of baryons
presented in table~\ref{table1}\ show
that the quantum treatment of the model {\it ab initio} improves
significantly the overlap with the experimentally-observed mass
spectrum when compared with results obtained using the standard rigid
rotator approach in the semiclassical version of the Skyrme model.
The individual quantum profile functions obtained can be used to 
calculate the magnetic moments and form factors of the baryons.

\paragraph{Acknowledgements.} 

The authors thank Paul Sutcliffe for valuable discussions and suggestions, 
and Niall MacKay for carefully reading the manuscript. 
VR also thanks the UK EPSRC for funding under grant EP/H000054/1.


\newcommand{\hepth}[1]{\href{http://arxiv.org/abs/hep-th/#1}{\tt hep-th/#1}}
\newcommand{\hepph}[1]{\href{http://arxiv.org/abs/hep-ph/#1}{\tt hep-ph/#1}}
\newcommand{\nuclth}[1]{\href{http://arxiv.org/abs/nucl-th/#1}{\tt hep-ph/#1}}
\newcommand{\arXiv}[1]{\href{http://arxiv.org/abs/#1}{\tt arXiv:#1}}



\begin{thebibliography}{}

\bibitem{Skyrme61}
    T.~H.~R.~Skyrme,
    Proc. Roy. Soc. \textbf{A 260}, 127 (1961).

\bibitem{Skyrme62}
    T.~H.~R.~Skyrme,
    Nucl. Phys. \textbf{31}, 556 (1962).

\bibitem{Adkins}
    G.~S.~Adkins, C.~R.~Nappi and E.~Witten,
    Nucl. Phys. \textbf{B 228}, 552 (1983).

\bibitem{Walliser}
    H.~Walliser,
    Nucl.~Phys.~\textbf{A 548}, 649 (1992).

\bibitem{Fujii87}
    K.~Fujii, A.~Kobushkin, K.~Sato and N.~Toyota,
    Phys.~Rev.~\textbf{D 35}, 1896 (1987).

\bibitem{Fujii88}
    K.~Fujii, K.~Sato and N.~Toyota,
    Phys.~Rev.~\textbf{D 37}, 3663 (1988).   
        
\bibitem{Bander}
    M.~Bander and F.~Hayot,
    Phys.~Rev.~\textbf{D 30}, 1837 (1984).
    
\bibitem{Braaten}
    E.~Braaten and J.~P.~Ralston,
    Phys.~Rev.~\textbf{D 31}, 598 (1985).

\bibitem{Acus98}
    A.~Acus, E.~Norvai\v{s}as and D.~O.~Riska,
    Phys.~Rev.~\textbf{C 57}, 2597 (1998), [\hepph{9712071}].

\bibitem{Jurciukonis}
    D.~Jur\v{c}iukonis, E.~Norvai\v{s}as, D.~O.~Riska,
    J.~Math.~Phys.~\textbf{46}, 072103 (2005), [\nuclth{0505003}].

\bibitem{Weigel}
    H.~Weigel,
    Chiral Soliton Models for Baryons, Springer, Berlin Heidelberg, 2008.


\bibitem{CHK}
    C.~G.~Callan, Jr., K.~Hornbostel and I.~R.~Klebanov,
    Phys.\ Lett.\ B {\bf 202}, 269 (1988).

\bibitem{NR}
    E.~Norvaisas and V.~Regelskis,
    Lith.\ J.\ Phys.\  {\bf 49}, 7 (2009)
    [\arXiv{0809.4157}].

\bibitem{Balachan}
    A.~P.~Balachandran, F.~Lizzi, V.~G.~J.~Rodgers,
    Nucl.~Phys.~\textbf{B 256}, 525 (1985).

\bibitem{Mazur}
    P.~O.~Mazur, M.~A.~Nowak, and M.~Praszalowicz,
    Phys.~Lett.~\textbf{B 147}, 137 (1984).

\bibitem{Sugano}
    R.~Sugano, Prog.~Theor.~Phys.~\textbf{46}, 297 (1971);
    T.~Kimura and R.~Sugano, \textit{ibid}. \textbf{47}, 1004 (1972);
    T.~Kimura, T.~Ohtani and R.~Sugano, \textit{ibid}. \textbf{48}, 1395 (1972);
    T.~Ohtani and R.~Sugano, \textit{ibid}. \textbf{50}, 1715 (1973).

\bibitem{Yabu}
    H.~Yabu and K.~Ando,
    Nucl.~Phys.~\textbf{B 301}, 601 (1988).
    
\bibitem{Park}
    N.~W.~Park, J.~Schechter and H.~Weigel,
    Phys.~Lett.~\textbf{B 228}, 420 (1989).    

\bibitem{Battye}
    R.~A.~Battye, S.~Krusch and P.~M.~Sutcliffe,
    Phys.~Lett.~\textbf{B 626}, 120 (2005), [\hepth{0507279}].


\end{thebibliography}
\end{document}